# Thermionic Current Beyond the Traditional Space Charge Limit Enabled by Trapped Ions in the Virtual Cathode

Z. L. Idema (zlidema@gmail.com) and M. D. Campanell (michaelcampanell@gmail.com)
*Lawrence Livermore National Laboratory, P.O. Box 808, Livermore, CA 94551, USA*

We show that ion trapping in virtual cathodes can raise the transmitted current of emitted electrons much closer to the full emission than is predicted by theories without trapped ions. The transmitted current is controlled by the well voltage whose value must adjust to balance the creation of low-energy ions within the well, and their leakage over the well. Our model quantifies these rates and derives the current in terms of system parameters for cathode emission into a plasma in several geometries. A general prediction is that the current as a function of emitted flux does not saturate at the traditional space charge limit (the onset of a well) but can reach far higher values. Improved performance might be achieved in plasma technologies with hot cathodes through suitable optimization of the trapped ion balance.

Studies of emitted electron space charge limits were vigorous a century ago [1] and remain so in modern research. A recent review of the topic by Zhang et al. [2] indicates that open questions remain due to the complexities of emission mechanisms, cathode geometries, time-dependent effects, quantum and relativistic effects, and use of plasma. But the underlying regulator of space charge current limitation is universal, a "virtual cathode" (VC) in a potential well that blocks some emitted electrons from escaping the surface. VC's are of high interest in diverse contexts including magnetic fusion [3], hypersonic vehicles [4], plasma processing arcs [5], dusty plasmas [6], Large Plasma Device experiments [7], satellite propulsion [8], plasma diagnostics [9], microelectronics [10], and laser-foil interactions (VC's cause the "recirculation" effect) [11].

Innovative efforts to overcome space charge limits such as pulsing the emission or the cathode biases have been considered [12]. Still, no known techniques yield major increases of average current over the traditional steady state limit corresponding to the onset of a VC [2]. This Letter will present theory showing for the first time that positive ion trapping in the VC well can considerably raise the steady state transmitted current in certain conditions. Such ions might be produced by photoionization [13], contact ionization at the cathode [14], or in the case to be treated here, when an upstream plasma feeds ions into the well.

Emission from cathodes into a plasma has been treated by many researchers since Langmuir's seminal 1929 paper [15]. Takamura's influential 2004 paper [16] features a rigorous model of the cathode sheath that set the stage for later variants [17]. Conventional works calculate the threshold emission of VC formation assuming the plasma ions propagate from the sheath edge to the cathode without collisions. Over the years, researchers have suggested that charge-exchange (CX) collisions of fast ions with cold neutrals may create trapped ions in VC wells and affect the sheath properties [18,19,20,21,22]. Tapped ion effects however have not yet been modeled self-consistently in cathode sheath physics models.

Intuition suggests that any extra positive space charge in the VC well from trapped ions should help offset the emitted electron space charge and enhance the current, as was argued by Lawless and Lam [21]. More recent studies indicate that trapped ion effects are more nuanced because their buildup leads to formation of a quasineutral plasma that has *zero net space charge* [23,24]. Campanell and Umansky showed that the trapped ion region, regardless of its size, cannot enhance the current from a cathode in pure planar geometry [23], but they argued that an enhancement might be possible in geometries where the trapped ions surround the cathode. Later, Johnson and Campanell confirmed such enhancement in simulations of a small square cathode in a large plasma [24]. Most recently, Yip *et al.* [25] published the first direct experimental study demonstrating that CX ion trapping causes current enhancement.

Current enhancement beyond the traditional space charge limit has profound implications in scientific research areas where surface-emitted electrons play a role. Until now, there is no model explaining physically how trapped ions enhance the current, how much enhancement is possible, and what factors control the degree of enhancement. Motivated by the need for theoretical understanding, this Letter will model the effects of trapped ions in the sheath of a plasma-facing cathode from first principles.

Let us consider a spherical strongly emitting cathode of radius $R_{cat}$ immersed in a large unmagnetized plasma with upstream density $N_{up}$. A schematic of the resulting cathode sheath is sketched in Fig. 1. A key concept illustrated in recent kinetic simulations [23,24] is that substantial trapped ion accumulation in the well creates an extended quasineutral region of nearly flat potential. (This outcome differs from when ions get orbitally trapped near an object while above its potential [26]). Within the "trapped ion plasma" (TIP), the trapped ions are neutralized by whatever thermoelectrons get past the potential barrier voltage $\Phi_{bar}$ at the cathode. The TIP will have a $\sim 1/r^2$ density profile in spherical geometry due to simple radial expansion of the emitted electron flow in a region with negligible electric field (we also neglect electron collisions in the TIP). We assume that the cathode bias $V_b$ is large relative to the temperature $T_e$ of plasma electrons ($q_e V_b \gg k_B T_e$) so that plasma electrons do not reach the TIP or the cathode.

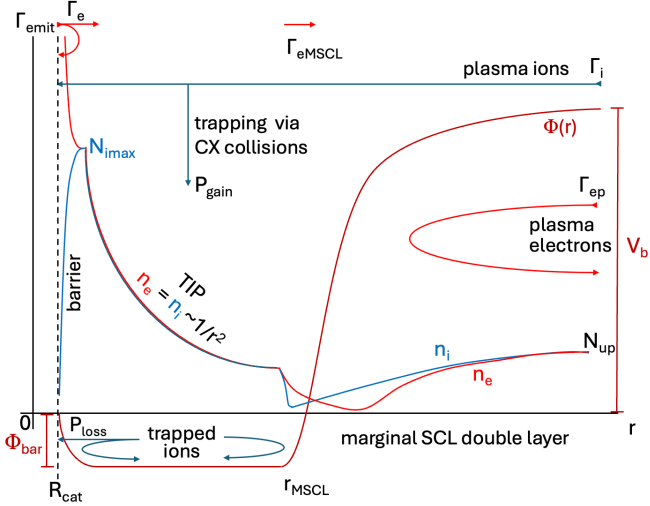

**FIG. 1.** Schematic of a thermionic cathode sheath with a quasineutral trapped ion plasma (TIP). Existence of TIP's has been demonstrated in fully kinetic simulation data in several geometries; see Fig. 4 of Ref. [23] and Figs. 2,4 of Ref. [24]. The present schematic and model equations treat a spherically symmetric cathode, but other geometries will be discussed. We note that the diagrammed flux densities $\Gamma$ (m$^{-2}$s$^{-1}$) are defined at the respective locations and are not radially conserved due to geometry.

Let $\Gamma_{emit}$ denote the flux density of thermoelectrons and $T_{emit}$ their temperature. Our goal is to calculate $\Gamma_e$, the portion that overcomes $\Phi_{bar}$ and produces current. Because the emitted electron flux density decays with distance from the cathode due to radial spreading, we emphasize that the $\Gamma_e$ calculated in this model is the net value measured by the cathode surface. Our model will also refer to the flux density of ions entering the double layer $\Gamma_i$ and the ambient flux density of bulk plasma electrons $\Gamma_{ep}$. These values are measured upstream and depend on plasma properties.

A foundational principle for cathode sheath physics with trapped ions is that the region between the upstream plasma and the TIP is a double layer (DL) with the familiar structure of a "marginal space charge limited" (MSCL) sheath. The MSCL current density $q_e\Gamma_{eMSCL}$ is the traditional limit without ion trapping. $\Gamma_{eMSCL}$ was calculated terms of system properties ($N_{up}$, $V_b$, etc.) by Takamura, *et al.* in planar geometry [16]. For tractability of calculation, we approximate that the DL region is thin enough to be almost planar so that the planar $\Gamma_{eMSCL}$ is emitted from the outer radial edge of the TIP, $r_{MSCL}$. One can see that trapped ions enhance the current beyond the MSCL limit by letting $\Gamma_{eMSCL}$ get emitted from a surface with radius $r_{MSCL} > R_{cat}$. We will estimate the current enhancement by patching the physics of the TIP to the MSCL sheath and the cathode. This approach has advantages over setting up a Poisson equation solution of the whole sheath region, which is a complex undertaking even without trapped ions [16,17].

Upstream ions are presumed to enter the (thin) DL from a Bohm presheath at the sound speed with flux density $\Gamma_i = N_{up}(k_BT_e/m_i)^{1/2}$ through a spherical shell of radius $\approx r_{MSCL}$. We presume that CX collisions occur with a mean free path $\lambda_{CX}$ that depends on cross sections and density of the neutral gas. The rate of trapped ion gain $P_{gain}$ is then equal to,

$$P_{gain} = (4\pi r_{MSCL}^2)\left(N_{up}\sqrt{\frac{k_BT_e}{m_i}}\right)\left(1 - e^{\frac{-(r_{MSCL}+l_{sides}-R_{cat})}{\lambda_{cx}}}\right). \quad (1)$$

The last factor in Eq. (1) is the probability of each passing ion suffering a CX collision while transiting the well radially. We neglected orbital motion [27] (angular velocity) effects of ions which may reduce the probability of reaching the cathode and raise the trapping rate. The $l_{sides}$ accounts for ion collection in the sides of the well bordering the flat-potential TIP. Although the sides are within the space charge regions presumed thin, we will see that their influence on $P_{gain}$ becomes significant if the TIP is also very thin, or nonexistent. An estimate for $l_{sides}$ was calculated in Eq. (2) by adapting Eq. (18) of Ref. [28]. Approximating planar geometry for the sides was reasonable since thermoelectrons have a thin local Debye length.

$$l_{sides} \approx \frac{2^{7/4}\varepsilon_0^{1/2}(k_BT_{emit})^{3/4}}{q_e(\pi m_e)^{1/4}\Gamma_e^{1/2}}arctan\left(\sqrt{\frac{\Gamma_{emit}}{\Gamma_e}-1}\right) \quad (2)$$

Without further collisions, trapped ions would oscillate indefinitely in the well as in an electrostatic trap [29]. We will assume that ions in the TIP have sufficient collisions with local neutrals to sustain a Maxwellian velocity distribution at a temperature $T_i$. The spatial maximum of the trapped ion density $N_{imax}$ is at the juncture of the TIP and emission barrier. Trapped ions leak over $\Phi_{bar}$ at a loss rate $P_{loss}$ estimated by Eq. (3). Kinetic corrections with error functions are avoided here so the equations reduce to a simpler, more transparent form.

$$P_{loss} = (4\pi R_{cat}^2)N_{imax}\sqrt{\frac{k_BT_i}{2\pi m_i}}e^{\frac{-q_e\Phi_{bar}}{k_BT_i}} \quad (3)$$

Invoking a common assumption that thermionic electrons are half-Maxwellian [16,27], those which overcome the barrier will produce current according to Eq. (4).

$$\Gamma_e = \Gamma_{emit}e^{\frac{-q_e\Phi_{bar}}{k_BT_{emit}}} \quad (4)$$

The barrier attenuates the spatial density of emitted electrons and their one-way flux density by the same factor. The emitted electron densities at the cathode $N_{emit}$ and the barrier edge (equal to $N_{imax}$ by TIP quasineutrality) thus carry the proportionality in Eq. (5). We neglected any *geometric* decay of flux density vs. radius within the thin barrier.

$$\frac{\Gamma_e}{N_{i,max}} = \frac{\Gamma_{emit}}{N_{emit}} = v_{emit} \equiv \sqrt{\frac{2k_BT_{emit}}{\pi m_e}} \quad (5)$$

The plasma density at the outer radial end of the TIP must be $\Gamma_{eMSCL}/v_{emit}$, so the $\sim 1/r^2$ TIP density profile requires,

$$N_{i,max} = \frac{\Gamma_{eMSCL}}{v_{emit}} \frac{r_{MSCL}^2}{R_{cat}^2}. \quad (6)$$

Using Eqs. (1-6), one can eliminate unknowns {$N_{imax}$, $r_{MSCL}$, $\Phi_{bar}$} to express $P_{gain}$ and $P_{loss}$ in terms of $\Gamma_e$ and $\Gamma_{emit}$. Setting $P_{gain} = P_{loss}$ provides an implicit steady state solution for $\Gamma_e$ versus $\Gamma_{emit}$. The result is Eq. (7) coupled to Eq. (2). Since we are interested in current enhancement beyond the traditional limit, $\Gamma_e$, $\Gamma_{emit}$, and $\Gamma_{ep}$ are expressed as primed quantities all normalized to $\Gamma_{eMSCL}$. Introduction of $\Gamma_{ep}$ absorbs the factors $N_{up}$ and $T_e^{1/2}$ after the ion mass cancels.

$$\frac{\Gamma'_e}{\Gamma'_{emit}} = \left[ \Gamma'_{ep} \sqrt{\frac{8\pi T_{emit}}{T_i}} \left( 1 - e^{-\frac{R_{cat}}{\lambda_{CX}}(\sqrt{\Gamma'_e} + \frac{l_{VC}}{R_{cat}} - 1)} \right) \right]^{T_i/T_{emit}} \quad (7)$$

A representative solution is shown in Fig. 2(a). The listed parameters are in a viable range for a low-temperature plasma containing a small hot cathode such as an emissive probe. We see there is an interval of $\Gamma_{emit}$' with two possible steady $\Gamma_e$' values. Fig. 2(b) shows the full space of the $P_{gain} - P_{loss}$ term as a function of $\Gamma_e$' and $\Gamma_{emit}$'. It indicates that the lower solution branch is unstable. Perturbations upward cause further accumulation of trapped ions and a transition to the upper branch which is stable. Perturbations downward cause a loss of all trapped ions and a transition to the MSCL current line ($\Gamma_e$' = 1). Fig. 2(c,d) indicate that the higher current enhancements coincide with weakening of $\Phi_{bar}$ towards zero and radial expansion of the TIP as expected.

The role of $l_{sides}$ is seen in Fig. 2(a) comparing solutions of (7) with and without $l_{sides}$. The solutions overlap under significant current enhancement because the TIP is wide compared to the well's sides. But when $\Gamma_e$' is near unity, the TIP is narrow, $r_{MSCL} \rightarrow R_{cat}$, and the ion trapping rate $P_{gain}$ in (2) is underestimated without $l_{sides}$. The lower solution branch without $l_{sides}$ approaches the abscissa but never intersects it, unphysically implying that as $\Gamma_{emit}$' $\rightarrow \infty$ there is a solution with no TIP despite an infinite well.

Fig. 2(b) has dashed lines outlining a hysteresis trajectory when varying the emission above the MSCL. If $\Gamma_{emit}$' is increased from unity, at first there is no solution. Physically, the potential well is not large enough in voltage and radius to confine enough ions to create a quasineutral TIP, so the expected outcome is $\Gamma_e$' = 1. Once $\Gamma_{emit}$' passes the abscissa-intercept of the lower solution branch with $l_{sides}$, a TIP forms and $\Gamma_e$' transitions to the upper branch. $\Gamma_e$' continues increasing with $\Gamma_{emit}$' up to the diagonal full emission line where $\Gamma_e$' = $\Gamma_{emit}$' and the "space charge effect" is nullified. From there if $\Gamma_{emit}$' is adjusted downward, $\Gamma_e$' will trace the stable branch and then drop back to unity.

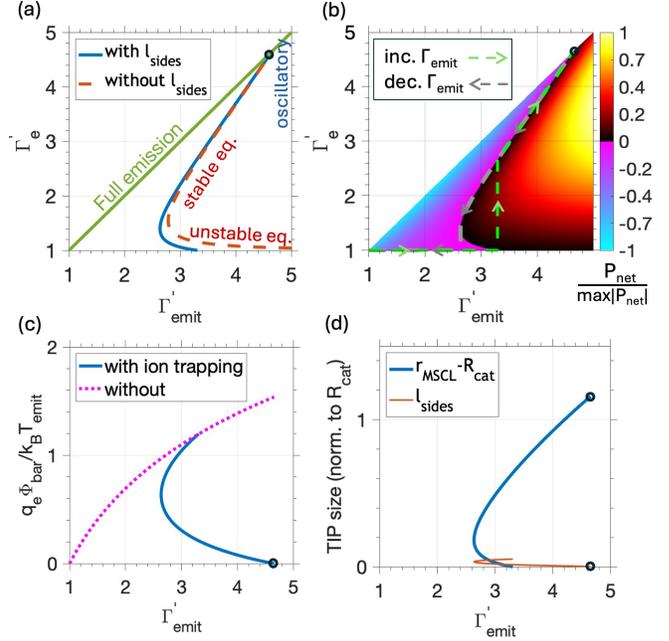

FIG. 2. Representative calculation of the trapped-ion current enhancement for a spherical hot cathode. Panel (a) plots solutions for $\Gamma_e$' versus $\Gamma_{emit}$' in equilibrium where $P_{net} \equiv P_{gain} - P_{loss} = 0$. Panel (b) estimates $P_{net}$ beyond the equilibria to explore solution stability. Panel (c) plots $\Phi_{bar}$ versus $\Gamma_{emit}$', comparing to conventional theory without ion trapping where $\Phi_{bar} \propto \ln(\Gamma_{emit}')$. Panel (d) illustrates the radial expansion of the TIP. Parameters set for this case include $T_e$ = 3eV, $V_b$ = 20V, $R_{cat}$ = 1mm, $T_{emit}$ = 0.2eV, $T_i$ = 0.08 eV, $\lambda_{CX}$ = 5mm, and $N_{up}$ = $10^{16}$m$^{-3}$.

What happens when $\Gamma_{emit}$' is raised beyond the crossing point of full emission? The expected outcome is an oscillatory regime where trapped ions gradually accumulate and get expelled by an explosive potential growth each time $N_{imax}$ crosses the maximum value that the thermoelectrons can neutralize ($N_{emit}$). This should result in low frequency sawtooth oscillations of current and could be the origin of unexplained oscillations measured in Fig. 1 of the experimental study in Ref. [30]. Further work analyzing TIP dynamics in the oscillatory regime is needed to answer whether the time-averaged $\Gamma_e$ levels off under increasing $\Gamma_{emit}$ or rises without a maximum. The maximum current transmittable in *steady state* is calculable in the present model by setting $\Gamma_{emit} = \Gamma_e$ and $l_{sides} = 0$ in Eq. (2),

$$\Gamma'_{e,max,spherical} = \left( 1 - \frac{\lambda_{CX}}{R_{cat}} \ln\left( 1 - \frac{1}{\Gamma'_{pe}} \sqrt{\frac{T_i}{8\pi T_{emit}}} \right) \right)^2. \quad (8)$$

The maximum for a cylindrical cathode is the square root of Eq. (8). The difference originates from the radii not being squared in the cylindrical variant of Eq. (6). The planar variant has a constant $N_{imax} = \Gamma_{eMSCL}/v_{emit}$. Consequently, $\Gamma_e$ is fixed at $\Gamma_{eMSCL}$, while $\Phi_{bar}$ and $P_{loss}$ are also inflexible. This theoretically answers why (a) no current enhancement is possible in pure planar geometry and (b) planar TIP's tend to exhibit runaway expansion, see Fig. 3 of Ref. [23].

Modelling TIP dynamics in systems with multidimensional cathode geometries, nonuniform emission, magnetic fields, and other complexities is left for future studies. We remark that because thermionic emission is isotropic in velocity space [31], the thermoelectrons from each individual point of a cathode in an unmagnetized plasma will spread radially as from a spherical emitter. It follows that a flat-potential TIP around a finite sized cathode of arbitrary shape, such as a flat disk, can expand towards a spherical shape making the resulting current enhancement share similar qualitative characteristics to the solutions here. The basic predictions of our theory, that ion trapping (a) causes formation of a TIP, (b) weakens $\Phi_{bar}$, and (c) brings $\Gamma_e$ closer to $\Gamma_{emit}$, are consistent with the fully kinetic simulations of a 2D square hot cathode in a plasma in Ref. [24]. A similar current enhancement could also be achieved in diodes without an "upstream plasma", if ions are produced local the cathode by other means.

Since trapped ions are the novel element of this work, we seek to understand how their parameters $\lambda_{CX}$ and $T_i$ affect current enhancement. Predictions can be made for a broad range of conditions by noting that in Eq. (7), $\Gamma_e'$ becomes a function of $\Theta_T \equiv T_i/T_{emit}$ and $H \equiv \lambda_{CX}/R_{cat}$ when the $l_{sides}$ influence is small; this is always applicable in the stable solution branch of primary interest assuming $l_{sides} < R_{cat}$ and $\Gamma_e'$ is not too near unity. Thus, Fig. 3 plots solutions without $l_{sides}$ as a function of $\Theta_T$ and $H$ so it can be used to understand how $R_{cat}$, $\lambda_{CX}$, $T_i$, and $T_{emit}$ affect the accessible stable currents. Other inputs are subsumed in $\Gamma_{ep}' \equiv \Gamma_{ep}/\Gamma_{eMSCL}$ of (7). We keep those from Fig. 2 because the influence and practical range of $\Gamma_{ep}'$ are narrow. For example, assuming $V_b$ is strong enough to confine plasma electrons, the calculations in Fig. 3 of Takamura's MSCL sheath model [16] show that $\Gamma_{eMSCL}$ only rises by a factor of ~2 for biases from ~4 to 100 times $k_B T_e q_e^{-1}$. The influence of $N_{up}$ cancels in $\Gamma_{ep}'$.

Since CX ions are born from neutrals, trapped ions may be as cold as an injected room-temperature neutral gas in low pressure devices [25], or as warm as $T_{emit}$ (tenths of eV) in high pressure devices such as arcs [5] where the particle species local to the cathode are in strong thermal contact. $\Theta_T$ values between 0.1 and 1.0 are therefore considered Fig. 3(a) for a fixed H. The maximum stable current is seen to increase at higher $\Theta_T$. The same is observed in cylindrical geometry in Fig. 3(b) but the enhancement is less and interestingly, the stable branch can vanish. The practical range of H can vary over orders of magnitude due to variabilities in cathode sizes, gas densities and CX cross sections. Fig. 3(c) shows that for fixed $\Theta_T = 0.2$, current enhancements are small for $H \leq 1$ and become substantial at higher H. These behaviors may be surprising. One could have expected from intuition that a faster ion trapping rate (small $\lambda_{CX}$, H) or slower loss rate (smaller $T_i$, $\Theta_T$) is needed to form a large TIP and access high currents.

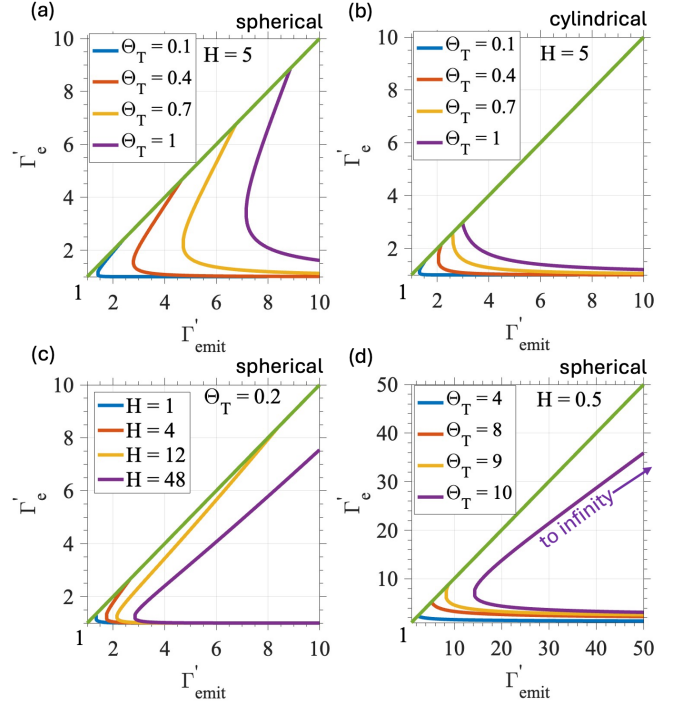

**FIG. 3.** Calculations showing the effects of geometry and variable system conditions on the current enhancement.

The opposite is observed because $P_{loss}$ has a fundamental maximum given by Eq. (3) with $\Phi_{bar} = 0$ and $N_{imax} = N_{emit}$. The maximum stable current sets in when the TIP is wide enough that $P_{gain}$ matches the maximum $P_{loss}$. Higher H or $\Theta_T$ values both allow the required gain-loss balance to be sustained at higher currents. At high enough $\Theta_T$ in Fig. 3(d), $\Gamma_e'$ no longer crosses the full emission line, so $\Gamma_e' \to \infty$ as $\Gamma_{emit}' \to \infty$. This profound "unbounded steady current" regime occurs when the argument of the logarithm in Eq. (8) becomes negative. Although it seems to require a less common scenario of $\Theta_T > 1$, meaning $T_i > T_{emit}$, that may be possible if supplemental energy gain of trapped ions takes place through instabilities [23] or heating. The unbounded behavior might be accessible for $\Theta_T < 1$ and at any H if trapped ion losses are enhanced by an auxiliary process not included here, such as volume recombination.

This Letter introduced the first hot cathode sheath model predicting an enhancement of current from trapped ions in the VC's potential well. The couplings between the cathode geometry, emission current and trapped ion dynamics revealed here should apply to a broad array of plasma systems with hot cathodes emitting beyond the traditional space charge limit. The degree of current enhancement was shown to be sensitive to the trapped ion gain and loss rates, and can rise without bound in ideal conditions. Recent experiments have revealed ways to actively increase both the trapped ion gain rate (e.g. by local puffing of neutrals to bolster CX collisions [25]) or loss rate (using dielectrics to draw out trapped ions near a cathode [19]). The improved theoretical understanding of trapped ion effects provided here combined with experimental controls could enable major advances in optimizing currents and mitigating space charge in technologies.

This work was performed under the auspices of the US Department of Energy by Lawrence Livermore National Laboratory under Contract No. DE-AC52-07NA27344, and supported by the US DOE Office of Science, Fusion Energy Sciences.